\documentclass[12pt]{article}
\usepackage{graphicx}
\usepackage{amsmath}
\usepackage{amssymb}
\usepackage{amsfonts}
\usepackage{float}
\usepackage{hyperref}
\usepackage{dsfont}  
\usepackage{slashed}  
\usepackage{booktabs}
\usepackage{multirow}
\usepackage{subfigure}
\newcommand{\beq}{\begin{eqnarray}}
\newcommand{\eeq}{\end{eqnarray}}
\usepackage{color}

\newcommand\pubdate{\today}

\def\Title#1{\begin{center} {\Large #1 } \end{center}}
\def\Author#1{\begin{center}{ \sc #1} \end{center}}
\def\Address#1{\begin{center}{ \it #1} \end{center}}

\newcommand\pubblock{\rightline{\begin{tabular}{l}  \\ 
         \pubdate  \end{tabular}}}
\newenvironment{Abstract}{\begin{quotation}  }{\end{quotation}}
\newenvironment{Presented}{\begin{quotation} \begin{center} 
             PRESENTED AT\end{center}\bigskip 
      \begin{center}\begin{large}}{\end{large}\end{center} \end{quotation}}

\begin{document}
\begin{titlepage}
 \pubblock
\vfill
\Title{Unraveling anomalies in Deeply Virtual Compton Scattering}
\vfill
\Author{Shohini Bhattacharya}
\Address{RIKEN BNL Research Center, Brookhaven National Laboratory, Upton, NY 11973, USA}
\Author{Yoshitaka Hatta}
\Address{Physics Department, Brookhaven National Laboratory, Upton, NY 11973, USA}
\Address{RIKEN BNL Research Center, Brookhaven National Laboratory, Upton, NY 11973, USA}
\Author{Werner Vogelsang}
\Address{Institute for Theoretical Physics, T\"{u}bingen University, Auf der Morgenstelle 14, 72076 T\"{u}bingen, Germany}

\vfill
\begin{Abstract}
We calculate the one-loop quark box diagrams relevant to polarized and unpolarized Deeply Virtual Compton Scattering by introducing an off-forward momentum $l^\mu$ as an infrared regulator. This  regularization approach allows us to reveal the poles associated with the chiral anomaly in the polarized scenario, as well as the trace anomaly in the unpolarized case. We provide an interpretation of our findings in the context of pertinent Generalized Parton Distributions (GPDs). Furthermore, we discuss the implications of these poles on the QCD factorization pertaining to Compton amplitudes.
\end{Abstract}
\vfill
\begin{Presented}
DIS2023: XXX International Workshop on Deep-Inelastic Scattering and
Related Subjects, \\
Michigan State University, USA, 27-31 March 2023 \\
     \includegraphics[width=9cm]{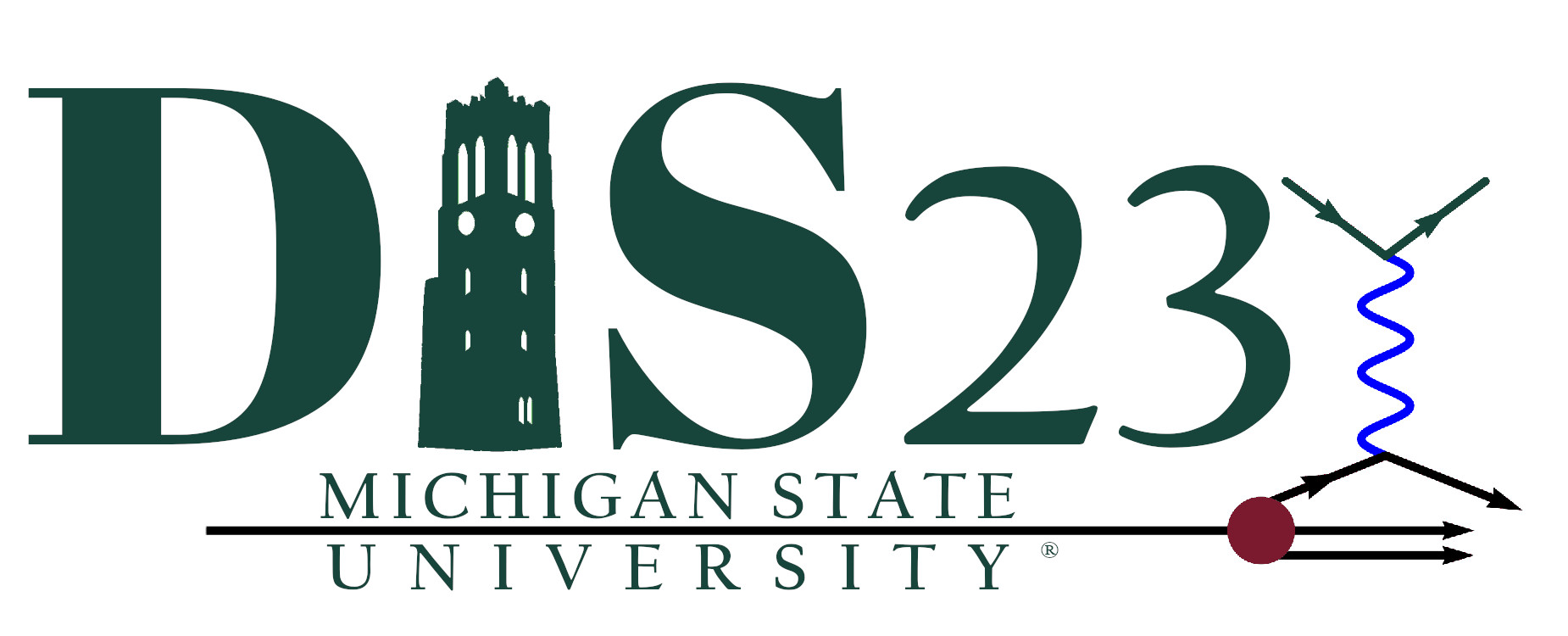}
\end{Presented}
\vfill
\end{titlepage}



\section{Introduction}
The historical narrative surrounding the axial anomaly's role in polarized Deep Inelastic Scattering (DIS) has twists and turns. Initially, during the late 1970s, it was utilized to constrain the one-loop gluonic correction to the initial moment of the singlet structure function $g_1(x)$. Subsequently, following the revelation of the ``spin crisis" by the European Muon Collaboration (EMC) in 1988, a suggestion emerged that the anomaly-induced gluon helicity $\Delta G$ might contribute to the intrinsic quark helicity $\Delta\tilde{\Sigma}$ in a way that addresses the unexpectedly small $\Delta \Sigma$ value~\cite{Carlitz:1988ab, Altarelli:1988nr}.

Theoretical objections to identifying the axial anomaly contribution as gluon helicity were evident from the start~\cite{Jaffe:1989jz,Bodwin:1989nz,Vogelsang:1990ug}. The gluonic contribution arises from the infrared realm of the triangle diagram featuring the Adler-Bell-Jackiw anomaly. Jaffe and Manohar strongly advocated for regulating the infrared singularity by computing it in off-forward kinematics~\cite{Jaffe:1989jz}. This method showcased the anomaly as a pole in momentum transfer $l$ in the matrix element of the singlet axial current $J_5^\mu$,
\begin{equation}
\langle p_2|J_5^\mu|p_1\rangle = \frac{n_f\alpha_s}{4\pi}\frac{il^\mu}{l^2}\langle p_2|F_a^{\alpha\beta}\tilde{F}_{\alpha\beta}^a|p_1\rangle \, .
\label{trian}
\end{equation}
This pole's appearance, along with the twist-four pseudoscalar operator $F\tilde{F}$, raised concerns about nonperturbative physics that eludes standard perturbative QCD. Notably, standard infrared regularization techniques, like dimensional regularization, sidestepped this problem in conventional forward kinematics.

The issue of the anomaly pole has once again come to the forefront with the recent work by Tarasov and Venugopalan, which serves to revitalize and reinforce Jaffe and Manohar's initial proposition (see above). Using the worldline formalism, they demonstrated that the box diagram exhibits a pole $1/l^2$ under off-forward kinematics~\cite{Tarasov:2021yll,Tarasov:2020cwl}. This represents a nonlocal extension (box) of the local relation (triangle) in Eq.~(\ref{trian}), unintegrated in the Bjorken variable $x$. 

\begin{figure}[t]
\centering
\includegraphics[width = 15cm]{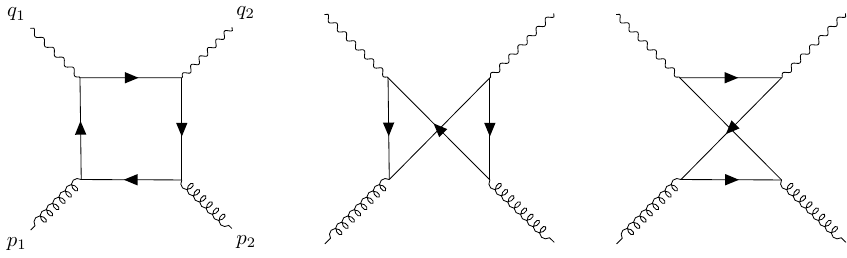}
\caption{Quark box diagrams for the Compton amplitude in  off-forward kinematics.}
\label{fig1}
\end{figure}

In these proceedings, we will explore the physics of anomaly poles in two directions. To begin, we will closely follow the methodology established in our previous works, Refs.~\cite{Bhattacharya:2022xxw, Bhattacharya:2023wvy}, to calculate the box diagram (see Fig.~\ref{fig1}) within the framework of standard perturbation theory under conditions of off-forward kinematics, i.e., $q_{1} = q + l/2, p_{1} = p - l/2, -q_{3}=q-l/2, -p_2 = p+l/2$ with $l^2 \equiv t \neq 0$. Specifically, we will focus on the computation of the ``Compton scattering amplitude", given by
\begin{equation}
T^{\mu\nu}=i\int \frac{d^4y}{2\pi}e^{iq\cdot y}\langle P_2|{\rm T}{J^\mu(y/2)J^\nu(-y/2)}|P_1\rangle =T^{\mu\nu}_{\rm sym} + iT^{\mu\nu}_{\rm asym} . \label{comp}
\end{equation}
Here, the electromagnetic current is denoted as $J^\mu=\sum_f e_f \bar{\psi}_f\gamma^\mu \psi_f$, where $f=u,d,s,..$ represents the flavor index. The subscript `asym/sym' refers to the antisymmetric/symmetric combination, corresponding to the polarized/unpolarized cases, of the photon polarization indices $\mu,\nu$. 
This calculation not only replicates the pole term found in \cite{Tarasov:2020cwl} but also yields typical perturbative corrections to the $g_1(x)$ structure function involving DGLAP splitting functions and a coefficient function (Sec.~\ref{sec:s1}). We elucidate these findings using GPDs $\tilde{H}$ and $\tilde{E}$. Although the pole's emergence seemingly poses challenges to QCD Compton amplitude factorization, we discuss how pole cancellations still warrant factorization. Secondly, we highlight that analogous poles arise in the unpolarized (symmetric)  portion of the off-forward Compton scattering amplitude $T^{\mu\nu}$ (Sec.~\ref{sec:s2}). This unpolarized pole, analogous to its polarized counterpart, relates to the trace anomaly. The off-forward photon matrix element of the energy-momentum tensor $\Theta^{\mu\nu}$ has a similar anomaly pole originating from the triangle diagram. The residue is linked to the matrix element of the twist-four scalar operator, characterizing the trace anomaly~\cite{Giannotti:2008cv,Armillis:2010qk}:
\begin{equation}
\langle p_2|\Theta^{\mu\nu}|p_1\rangle \sim \frac{1}{l^2}\langle p_2|F^{\alpha\beta}F_{\alpha\beta}|p_1\rangle \, . \label{traceano}
\end{equation}
We derive the unintegrated version of this result and connect it to unpolarized GPDs $H$ and $E$. 

\section{Antisymmetric part of Compton amplitude}
\label{sec:s1}

In the antisymmetric case we define 
\beq
{\cal J}^\alpha&\equiv&-\epsilon^{\alpha\beta\mu\nu}P_\beta {\rm Im}T^{\rm asym}_{\mu\nu} \, .
\eeq
(In the limit where $l^\mu \to 0$, the expression ${\cal J}^\alpha=g_1(x_B)\bar{u}(P)\gamma^\alpha\gamma_5 u(P)=2g_1(x_B)S^\alpha$ becomes a direct proportion to the polarized DIS $g_1$ structure function, considering the forward kinematics.) Employing the Mathematica package `Package-X'~\cite{Patel:2015tea}, we 
evaluated the box diagrams in the region where $\Lambda^2_{\rm QCD} \ll l^2 \ll Q^2$. For massless quarks circulating in the loop, the resulting expression is as follows:
\beq
{\cal J}^\alpha |_{\rm box} \approx \frac{1}{2} \frac{\alpha_s}{2\pi}\left(\sum_f e_f^2\right)  \bar{u}(P_2)\Biggl[\left(\Delta P_{qg}\ln\frac{Q^2}{-l^2}+\delta C_g^{\rm off}\right)\otimes \Delta G(x_B) \gamma^\alpha\gamma_5  \nonumber \\ 
 + \frac{l^\alpha}{l^2} \delta C_g^{\rm anom} \otimes \tilde{{\cal F}}(x_B) \gamma_5 \Biggr]u(P_1)\, . \label{off}
\eeq
Here, we introduce the notation $A\otimes B(x_B) \equiv \int_{x_B}^1 \frac{dx}{x} A\left(\frac{x_B}{x}\right)B(x)$. (In equation (\ref{off}), we disregard terms that vanish as $l\to 0$, inclusive of the $\xi$-dependent coefficients. For a complete result with the $\xi$-dependence, we refer to Ref.~\cite{Bhattacharya:2023wvy}.) The polarized $g\to q$ splitting function $\Delta P_{qg}(\hat{x})=2T_R(2\hat{x}-1)$ is defined, where $T_R=\frac{1}{2}$ is the color factor.
\beq
\delta C_g^{\rm off}(\hat{x}) = 2T_R(2\hat{x}-1)\left( \ln \frac{1}{\hat{x}(1-\hat{x})}-1\right)
\eeq
is the coefficient function. $\tilde{{\cal F}}(x)$ pertains to the twist-four pseudoscalar parton distribution as discussed in references~\cite{Tarasov:2020cwl,Hatta:2020ltd, Radyushkin:2022qvt},
\beq
\tilde{{\cal F}}(x,l^2)&\equiv & \frac{iP^+}{\bar{u}(P_2)\gamma_5u(P_1)}\int \frac{dz^-}{2\pi} e^{ixP^+z^-} \langle P_2|F_a^{\mu\nu}(-z^-/2)\tilde{F}^a_{\mu\nu}(z^-/2)|P_1\rangle \, ,
\eeq
entering Eq.~(\ref{off}) with the anomalous coefficient function 
\beq 
\delta C_g^{\rm anom}(\hat{x}) =4T_R(1-\hat{x}) \, .
\eeq
The pole term above is consistent with the finding in \cite{Tarasov:2020cwl}. In the usual DGLAP splitting term, we see that the collinear singularity is regularized by $l^2$. 

We then proceed to interpret this outcome in the context of the GPDs $\tilde{H}$ and $\tilde{E}$ by establishing a correspondence between Eq.~(\ref{off}) and the QCD factorization expression for the Compton amplitude. It's important to highlight that due to the distinct Lorentz structure proportional to $l^\alpha$ in the pole term, the pole is exclusively connected with the GPD $\tilde{E}$ and not $\tilde{H}$. Despite the emergence of this pole (indicating sensitivity to infrared physics) which challenges factorization, we contend that upon calculating perturbative corrections to quark GPDs within a gluon target, we observe an identical pole structure in $\tilde{E}$ itself (see Ref.~\cite{Bhattacharya:2023wvy}). Therefore, we establish that this pole is an integral part of the GPD itself and must be absorbed within the GPD. This reasoning serves as a justification for the validity of factorization.

Our work thus highlights a new perspective on GPD: the 
physics associated with the axial anomaly is inherently present within the GPD $\tilde{E}$. (This understanding arises from the fact that the residue of this pole corresponds to $\tilde{{\cal F}}\sim \langle  F \tilde{F}\rangle$.) This insight is extensively explored in Refs.~\cite{Bhattacharya:2022xxw,Bhattacharya:2023wvy}, where we have also postulated that this anomaly-related pole should be  counteracted by another massless pole so  that there is no massless pole in the GPD $\tilde{E}$, as it should. Such a pole cancellation mechanism is known in the context of the flavor-singlet pseudscalar axial form factor $F_P(t)$ and is responsible for the  nonperturbative mass generation of the $\eta'$ meson \cite{Veneziano:1979ec}. The same scenario is realized at the GPD level because $F_P(t)\sim \int dx \tilde{E}(x,t)$.

\section{Symmetric part of Compton amplitude}
\label{sec:s2}
Interestingly, the $1/l^2$ pole also surfaces within the symmetric component (in $\mu\nu$) of the Compton amplitude. As explained in preceding sections and further detailed below, this pole's appearance in this context is intricately linked to the QCD trace anomaly.

Our computational results have unveiled a tensor structure that is notably more intricate than the conventional one:
\beq
& {\rm Im} T^{\mu\nu}_{\rm sym}|_{\rm box}= 
\left(-g^{\mu\nu}+\frac{q^\mu q^\nu}{q^2}\right) F^{\rm off}_1(x_B,l) + \left(P^\mu-\frac{P\cdot q}{q^2}q^\mu\right)\left(P^\nu-\frac{P\cdot q}{q^2}q^\nu\right)\frac{2x_BF^{\rm off}_2(x_B,l)}{Q^2} 
\nonumber \\
& +\dots .
\label{offf}
\eeq
Here, the ellipses encompass terms associated with tensors such as $l^{(\mu} P^{\nu)}$, $l^{(\mu} q^{\nu)}$, and $l^\mu l^\nu$, designed in a manner that adheres to the conservation law $q_\mu {\rm Im}T^{\mu\nu}_{\rm sym}=0$. The structure functions $F_{1,2}^{\rm off}$ can be related to the unpolarized GPDs $H, E$~\cite{Bhattacharya:2022xxw}. Below we maintain the complete dependence on the skewness parameter $\xi$. To achieve this, we perform the substitution $l^\mu \approx -2\xi P^\mu \approx - 2\hat{\xi}p^\mu$ within the tensors $l^{(\mu} P^{\nu)}$, $l^{(\mu} q^{\nu)}$, and $l^\mu l^\nu$, while refraining from applying this substitution to potential pole terms $\propto 1/l^2$. This approach leads us to a simplified representation of Eq.~(\ref{offf}): 
\beq
 {\rm Im} T^{\mu\nu}_{\rm sym} |_{\rm box} &\approx& 
\left(-g^{\mu\nu}+\frac{q^\mu q^\nu}{q^2}\right) \bar{F}^{\rm off}_1(x_B,l) \nonumber \\
&&+ \left(P^\mu-\frac{P\cdot q}{q^2}q^\mu\right)\left(P^\nu-\frac{P\cdot q}{q^2}q^\nu\right) \frac{2x_B\bar{F}^{\rm off}_2(x_B,l)}{Q^2} \, ,
\eeq
where (setting $\xi=0$ in non-pole terms)
\beq
\bar{F}^{\rm off}_1(x_B,l) \approx \frac{1}{2} \frac{\alpha_s}{2\pi}\left(\sum_f e_f^2\right)  \bigg[ \left(P_{qg}\ln \frac{Q^2}{-l^2}+C^{\rm off}_{1g}\right) \otimes g(x_B)  \nonumber \\
+ \frac{1}{l^2}C^{\rm anom} \otimes' {\cal F} (x_B,\xi,l^2)\frac{\bar{u}(P_2)u(P_1)}{2M}\bigg] \, ,
\label{offsym} 
\eeq
and similarly for $\bar{F}^{\rm off}_2(x_B,l)$. We identify the anticipated structure of the one-loop corrections linked to the unpolarized gluon PDF $g(x)$, featuring the splitting function $P_{qg}(\hat{x})=2T_R((1-\hat{x})^2+\hat{x}^2)$. The pertinent coefficient function is expressed as follows: 
\beq
\begin{split}
& C_{1g}^{\rm off}(\hat{x}) =2T_R ((1-\hat{x})^2+\hat{x}^2) \left(\ln \frac{1}{\hat{x}(1-\hat{x})}-1\right) \, . \label{coeff}
\end{split}
\eeq
Furthermore, we ascertain the presence of a $1/l^2$ pole in $\bar{F}_1^{\rm off}$ (and $\bar{F}_2^{\rm off}$). In Eq.~(\ref{offsym}), the convolution formula is expressed as follows:
\beq
C^{\rm anom}\otimes' {\cal F} (x_B,\xi,l^2) \equiv \int_{x_B}^1\frac{dx}{x}K(\hat{x},\hat{\xi}){\cal F} (x,\xi,l^2) \nonumber \\
-\frac{\theta(\xi-x_B)}{2}\int_{-1}^1 \frac{dx}{x}L(\hat{x},\hat{\xi}){\cal F} (x,\xi,l^2) \, , \label{two}
\eeq
where 
\beq
&&K (\hat{x},\hat{\xi})  
=2T_R\frac{\hat{x}(1-\hat{x})}{1-\hat{\xi}^2} \, , \qquad L(\hat{x},\hat{\xi}) = 2T_R\frac{\hat{x}(\hat{\xi}-\hat{x})}{1-\hat{\xi}^2} \, . \label{whitec}
\eeq
The twist-four scalar gluon  GPD is defined as \cite{Hatta:2020iin,Radyushkin:2021fel}
\beq
{\cal F} (x,\xi,l^2)&=&-4xP^+M \int \frac{dz^-}{2\pi} e^{ixP^+z^-} \frac{ \langle P_2|F^{\mu\nu}(-z^-/2)F_{\mu\nu}(z^-/2)|P_1\rangle}{\bar{u}(P_2)u(P_1)} \, . \label{four}
\eeq

Moving forward, similar to Sec.~\ref{sec:s1}, we proceed to elucidate the implications of this outcome within the framework of the GPDs $H$ and $E$ by establishing a correspondence between Eq.~(\ref{offsym}) and the QCD factorization expression for the Compton amplitude. Notably, a distinction arises: while in the polarized context, the pole's association remains exclusive to $\tilde{E}$, in the unpolarized scenario, the pole connects with both GPDs, $H$ and $E$. Despite the emergence of this pole, indicative of its sensitivity to infrared physics and thereby posing a challenge to factorization, we demonstrated that upon calculating perturbative corrections to quark GPDs within a gluon target, a mirrored pole structure in $H$ and $E$ surfaces (as detailed in Ref.~\cite{Bhattacharya:2023wvy}). This compellingly establishes the pole as an integral component of the GPD itself. After absorbing the pole into the GPD, we reaffirm the validity of factorization.

Our work once again brings to light a significant insight: the physics linked to the trace anomaly and therefore the origin of hadron masses is inherently embedded within the GPDs $H, E$. (This understanding emerges due to the residue of this pole being ${\cal F} \sim\langle FF\rangle$.) As 
explored in Refs.~\cite{Bhattacharya:2022xxw,Bhattacharya:2023wvy}, we further postulate that this anomaly-related pole must be cancelled nonperturbatively such that the GPDs $H,E$, as well as their moments, the gravitational form factors, do not have a massless pole. The  precise mechanism behind this scenario pertains to the nonpertubative mass generation of glueballs and will be explored in future.

\section{Summary}
In these proceedings, we have discussed the factorization of Compton amplitudes within the $\Lambda^2_{\text{QCD}} \ll t \ll Q^2$ regime. After absorbing infrared singularities into GPDs, we have actually recovered the standard factorization formula with the same coefficient functions as in the $\overline{\rm MS}$ scheme (see \cite{Bhattacharya:2023wvy} for the details). This means that the result can be smoothly connected to the regime $t\sim \Lambda_{\text{QCD}}^2$. Moreover, by employing $t$ as a regulator, we have uncovered the intricate connection to the axial and trace anomalies of QCD, underscoring their profound significance as nonperturbative underpinnings for GPDs. This connection aligns with the established restrictions that these anomalies impose on the moments of GPDs, the axial and gravitational form factors. This significant development not only extends the boundaries of GPD research to encompass quantum anomalies, but also opens up new avenues for investigating their manifestations in both high-energy exclusive processes as well as the realm of Lattice QCD studies.

\section*{Acknowledgements}
S.~B. and Y.~H. are supported by the U.S. Department of Energy under Contract No. DE-SC0012704, and   Laboratory Directed Research and Development (LDRD) funds from Brookhaven Science Associates. Y.~H. is also supported by the framework of the Saturated Glue (SURGE) Topical Theory Collaboration. W.~V. has been supported by Deutsche Forschungsgemeinschaft (DFG) through the Research Unit FOR 2926 (project 409651613). 


\bibliographystyle{unsrt}
\bibliography{references.bib}

\end{document}